MDPI

*Article*

# An Ambient Intelligence-Based Human Behavior Monitoring Framework for Ubiquitous Environments

**Nirmalya Thakur** * and Chia Y. Han

Department of Electrical Engineering and Computer Science, University of Cincinnati, Cincinnati, OH 45221-0030, USA; han@ucmail.uc.edu
* Correspondence: thakurna@mail.uc.edu

**Abstract:** This framework for human behavior monitoring aims to take a holistic approach to study, track, monitor, and analyze human behavior during activities of daily living (ADLs). The framework consists of two novel functionalities. First, it can perform the semantic analysis of user interactions on the diverse contextual parameters during ADLs to identify a list of distinct behavioral patterns associated with different complex activities. Second, it consists of an intelligent decision-making algorithm that can analyze these behavioral patterns and their relationships with the dynamic contextual and spatial features of the environment to detect any anomalies in user behavior that could constitute an emergency. These functionalities of this interdisciplinary framework were developed by integrating the latest advancements and technologies in human–computer interaction, machine learning, Internet of Things, pattern recognition, and ubiquitous computing. The framework was evaluated on a dataset of ADLs, and the performance accuracies of these two functionalities were found to be 76.71% and 83.87%, respectively. The presented and discussed results uphold the relevance and immense potential of this framework to contribute towards improving the quality of life and assisted living of the aging population in the future of Internet of Things (IoT)-based ubiquitous living environments, e.g., smart homes.

## 1. Introduction

The elderly population across the globe is increasing at a very fast rate. It has been estimated [1] that by the year 2050, around 20% of the world's population will be aged 60 years or more. Aging is associated with several issues and limitations that affect a person's quality of life. According to [2], in the United States, approximately 8 out of every 10 elderly people have some form of chronic diseases, and approximately 5.4 million older adults have Alzheimer's. People living longer are causing a significant increase in the old-age dependency ratio, which is the ratio of the count of elderly people to that of the working population. On a global scale, this ratio is expected to increase from 11.7% to 25.4% over the next few years [2]. In addition to this, the population of elderly people with ages 80 and above is expected to triple within the next few years [3]. This increase in the population of older adults would bring several sociological and economic needs to the already existing challenges associated with aging. This constantly increasing elderly population is expected to impact society in multiple ways, as outlined below [3]:

i. A rise in cost of healthcare: at present, the treatment of older adults' accounts for 40% of the total healthcare costs in the United States even though older adults account for around 13% of the total population.

ii. Diseases affecting greater percentage of the population: with the increasing elderly population, there will be an increased number of people with diseases like Parkinson's and Alzheimer's, for which there is yet to be a proper and definitive cure.

iii.     Decreased caregiver population: the rate of increase of caregivers is not as high as the increasing rate of the elderly population.

iv.     Quality of caregiving: caregivers would be required to look after multiple older adults, and quite often they might not have the time, patience, or energy to meet the expectations of caregiving or to address the emotional needs of the elderly.

v.     Dependency needs: with multiple physical, emotional, and cognitive issues associated with aging, a significant percentage of the elderly population would be unable to live independently.

vi.     Societal impact: the need for the development of assisted living and nursing facilities to address healthcare-related needs.

With the decreasing count of caregivers, it is necessary that the future of technology-based living environments, e.g., smart homes and smart cities use technology-based services to address these needs and create assisted living experiences for the elderly. Over the last few years, researchers [4] have focused on developing assistive systems and devices according to a new paradigm, "ambient intelligence." Ambient intelligence may broadly be defined as a computing paradigm that uses information technology and its applications to enhance user abilities and performance through interconnected systems that can sense, anticipate, adapt, predict, and respond to human behavior and needs.

Human behavior is associated with performing activities in various environments and settings. An activity may broadly be defined as an interaction between a subject and an object for the subject to achieve a desired end goal or objective. This is typically represented as "S <-> O," where S stands for the 'subject' and O stands for the 'object.' Here, the subject is the user or the individual performing the activity, and the objects can be one or more context parameters present in the confines of the user's spatial orientation that are a part of the activity. To complete any given activity, the subject performs a set of related and sequential tasks or actions on one or more objects that depends on the kind of activity to be performed. These tasks or actions, along with their associated characteristic features, represent the user interactions related to the specific activity [5].

There can be various kinds of activities that a user performs in different environments with different spatial configurations. Activities that are crucial to one's sustenance and are performed within the confines of one's living space, e.g., personal hygiene, dressing, eating, maintaining continence, and mobility, are collectively termed as activities of daily living (ADLs) [6]. Based on the interaction patterns of the subject and object during activities, there are five broad characteristics of ADLs—(1) sequential, (2) concurrent, (3) interleaved, (4) false start, and (5) social interactions [5]. When multiple ADLs occur either at the same time or in a sequence or in a combination, they may exhibit more than one of these characteristics. Figure 1 shows four typical scenarios of different ADLs—A1, A2, A3, and A4—that can occur, where a number of these characteristics were exhibited by the activity sequences and combinations.

Elderly people need assistance to carry out ADLs due to the various bodily limitations and disabilities that they face with aging. An important aspect towards creating assisted living experiences in smart homes for the aging population is to monitor their interactions with their surroundings during ADLs [7]. The semantic analysis of user interactions during any ADL involves the monitoring of the associated behavioral patterns with respect to contextual, spatial, and temporal information. This analysis helps in interpretation of user performance during ADLs, as well as allowing for the detection of any anomalies that could constitute an emergency. For example, a person lying on a bed in a bedroom for several hours at night would mean that the person is taking rest, but if the same activity of lying is tracked to be taking place at the bathroom at the same time, it could mean an emergency situation resulting from a fall or unconsciousness, which needs the attention of caregivers or medical practitioners. In addition to aiding during ADLs, human behavior monitoring allows for the early detection of various forms of cognitive impairment, dementia, Alzheimer's, and a range of other limitations associated with old age [8]. Since it is not practically possible to manually access an older adult's behavior, it is the need of the hour to develop technology-based solutions with ambient intelligence to address this challenge. This served as the main motivation for the development of this framework that leverages the potential at the intersection of multiple disciplines including human–computer interaction, the Internet of Things (IoT), ubiquitous computing, machine learning, and pattern recognition.

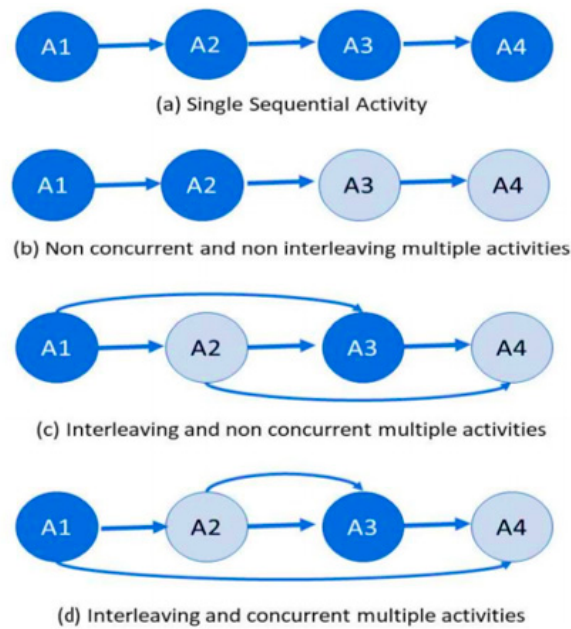

**Figure 1.** Representation of four typical scenarios of different activities of daily living (ADLs)—(**a**) A1, (**b**) A2, (**c**) A3, and (**d**) A4—that can occur, where different characteristics of ADLs are exhibited by the activity sequences and combinations.

To summarize, the scientific contributions of this paper are as follows:

1.  It provides a novel approach to perform the semantic analysis of user interactions on the diverse contextual parameters during ADLs in order to identify a list of distinct behavioral patterns associated with different complex activities performed in an IoT-based environment. These behavioral patterns include walking, sleeping, sitting, and lying. This functionality was developed and implemented by using a k-nearest neighbor algorithm (k-NN) classifier. The performance accuracy of this approach was found to be 76.71% when it was evaluated on a dataset of ADLs.
2.  It provides a novel intelligent decision-making algorithm that can analyze such distinct behavioral patterns associated with different complex activities and their relationships with the dynamic contextual and spatial features of the environment in order to detect any anomalies in user behavior that could constitute an emergency, such as a fall or unconsciousness. This algorithm was developed and implemented by using a k-NN classifier, and it achieved an overall performance accuracy of 83.87% when tested on a dataset of ADLs.

This paper is organized as follows. We present an overview of the related works in Section 2. The proposed framework is introduced and explained in Section 3. Section 4 discusses the results and performance characteristics of this framework. In Section 5, we discuss the limitations and drawbacks in the existing works and outline how our framework addresses these challenges and outperforms these existing systems in terms of their technical characteristics, functionalities, and operational features. It is followed by Section 6, where the conclusion and scope for future work are outlined.

## 2. Literature Review

This section outlines the recent works in the fields of human behavior research, i.e., assistive technology, Internet of Things, human–computer interaction, and their related disciplines for creating assisted living experiences in the future of technology-laden living environments, e.g., smart homes and smart cities.

A system comprised of wireless sensors to track and interpret human motion data for performing activity recognition was proposed by Azkune et al. [9]. The system consisted of an approach of activity clusters that were developed by using knowledge engineering principles. Based on these clusters, the associated patterns of human motion related to different activities could be tracked and interpreted by this system. Boualia et al. [10] proposed a Red Green Blue (RGB) frame analysis-based activity recognition framework with a specific focus on the study of human poses during different activities. The framework used a Convolutional Neural Network (ConvNet) architecture that was adapted for a regression problem and a support vector machine (SVM) classifier to detect activities. The authors evaluated the performance characteristics of their framework by testing it on two activity recognition datasets. Kasteren et al. [11] proposed a hidden Markov model-based architecture that analyzed the multimodal characteristics of sensor data for activity recognition. The authors used a recorded dataset and developed its annotation by using an off-the-shelf sensor, the Jabra BT250v. The Jabra BT250v was used to develop annotations for all the activities performed during each day, and these annotations were then used to train the hidden Markov model-based architecture for activity recognition. Cheng et al. [12] developed a framework that used concepts from computer vision, image processing, and video-data analysis to track and detect activities for both one and multiple users in the confines of a given IoT-based space. The approach combined characteristic features of motion data and user appearance information, as well as the spatiotemporal features of user behavior to train multiple learning models. The authors evaluated their approach by testing it on a dataset of activities. Skocir et al. [13] developed an artificial neural network-driven architecture that tracked human motion during different activities, with a specific focus on detecting enter and exit events in the confines of a given spatial environment, e.g., entering and exiting a room. The architecture used two IoT-based sensors with distinct functionalities to develop its foundation. One of these sensors was used to detect the presence or absence of the user, and the other sensor was used to detect whether the door was opened or closed. A dataset of different activities was used by the authors to test and discuss the performance characteristics of their approach.

The work done by Doryab et al. [14] involved the development of a task recommendation system to augment performances of medical practitioners in hospitals. This recommendation system was sensor technology-driven and focused on recommending tasks specifically related to different kinds of surgeries. The sensor data were used to detect the current action being performed by the user, and based on the same action, tasks associated with co-located activities were recommended by the system. A sensor network-driven activity assistance framework with the aim to assist users to perform different kinds of activities was proposed by Abascal et al. [15]. This work was specifically focused on helping elderly people with different kinds of impairments such as sensory, motor, or cognitive. In addition to performance characteristics, the authors also evaluated the accessibility, usability, and validity of their system. A system for the behavior monitoring of older adults in smart homes that used concepts of activity recognition and analysis was proposed by Chan et al. [16]. This system collected human motion data related to specific ADLs—walking, sleeping, and using the bathroom. The authors conducted real-time experiments in an Alzheimer's unit with a specific focus on studying and analyzing the human behavior and activities of people with Alzheimer's. Rashid et al. [17] developed a wearable neckband for human eating activity recognition and analysis. The system had a functionality to automatically update its database to adjust depending on the changing eating styles and eating habits of users. It used an artificial neural network-based approach that could detect four eating states—chewing, swallowing, talking, and idle. Siraj et al. [18]

developed a framework to recognize small activities, such as cooking, that are performed with other complex activities during a day. The authors used multiple machine learning models including those of deep learning, convolutional neural network, and gated recurrent unit to train their framework for the recognition of tasks and actions associated with the activity of cooking. They evaluated their framework on a dataset which consisted of various actions and tasks related to cooking. Mishra et al. [19] proposed a system that used video data for activity recognition and analysis. The system consisted of a spatiotemporal approach defined by considering the fuzzy lattices of the video frames. These lattices were described by kinetic energy, which was calculated by the Schrödinger wave equation. The system could detect any changes in human behavior or motion based on the change in kinetic energy associated with these lattices. In [20], Fu et al. proposed a wireless wearable sensor-driven device that could perform activity recognition. The device consisted of an air pressure sensor and an inertial measurement unit to study and analyze human behavior related to different activities. The wearable used a transfer learning approach to perform personalized activity recognition. The work done by Yared et al. [21] involved the development of an intelligent activity analysis framework to reduce accidents in the kitchen area. The authors analyzed multiple activities performed in the kitchen to identify characteristic features such as gas concentration, smoke, the temperature of utensils, and the temperature of burner that needed to be monitored to detect any accidents. The findings of this work listed a set of factors that were responsible for most kitchen accidents. Angelini et al. [22] developed a smart bracelet that could collect multiple features of a user's movement data to interpret the health status of the user. It also had the functionality to remind the user of their routine medications. The bracelet was developed to work for different kinds of indoor and outdoor activities. The authors conducted usability studies to discuss the effectiveness of this bracelet.

In the work done by Dai et al. [23], the dynamics of the motion data coming from the user's android phone were analyzed to detect falls. The authors developed a proof-of-concept model that was based on an Android phone that collected real-time behavior-related data of the user. The architecture of the system was developed in a specific way to ensure that it did not contribute to high central processing unit (CPU) usage and did not occupy a significant percentage of the computer's random-access memory (RAM). The results discussed by the authors showed that the average CPU usage was 7.41% by the system, and it occupied about 600 KB on the RAM. Kong et al. [24] proposed a depth recognition and distance-based algorithm for detecting falls. The algorithm tracked the distance between the neck of the user and the ground, and if the distance was found to decrease with a situation lasting greater than a minute, then the algorithm interpreted the situation as a fall. Shao et al. [25] proposed an approach that analyzed the characteristics of floor vibrations to detect falls. The authors performed experiments with objects and humans falling on the ground to study the characteristics of floor vibrations. The system consisted of a k-means classification approach to detect falls. Chou et al. [26] proposed an Electrocardiography (ECG)-based system for fall detection. The system consisted of a smart cane with an ECG detection circuit along with other sensors to study the behavioral patterns of the user. The authors developed and implemented a microcontroller-based circuit that could detect falls based on the data collected from the ECG circuit and the associated sensors. In [27], Keaton et al. proposed an WiFi channel state-based approach for the detection of falls in IoT-based environments. The authors developed a neural network-based learning model that could study, track, and analyze the changes in WiFi channel state based on normal behaviors and falls. Anceschi et al. [28] proposed a machine learning-based wearable system for fall detection in a workplace environment. To develop and train the machine learning model, the authors merged four different datasets that consisted of diverse activities performed in a workplace. This device used a couple of IoT-based off-the-shelf products that worked in coordination with a microcontroller circuit to detect falls from human motion data. Mousavi et al. [29] used acceleration data available from smartphones to develop a fall detection system. This system consisted of an SVM classifier that interacted with the triaxial accelerometer data coming from a smartphone that had an IOS operating system. The system also had a feature to alert caregivers via either an SMS or email when a fall was detected.

Despite these recent advances in this field, there are still several limitations and challenges. For instance, (1) a number of these works have superficially focused on activity recognition without an analysis of the fine grain characteristics of activities and the associated dynamics of human behavior; (2) several activity analysis approaches are confined to specific tasks and cannot always be applied seamlessly to other activities; (3) a number of these methodologies have been developed and implemented for specific settings with a fixed set of context parameters and environment variables, and their real world deployment is difficult because the real world environments are different compared to such settings; (4) the video-based systems may have several challenges related to the categorization and transcription of data, the selection of relevant fragments, the selection of camera angle, and the determination of the number of frames; (5) some of the fall detection technologies are built for specific operating systems, devices, or gadgets and cannot be implemented on other platforms; (6) some of these systems have a dependency on external parameters, such as floor vibrations, that can affect the readings and performance characteristics; and (7) some of the systems are affected by user diversity such as the user's height and weight. To add to the above, some of these works have focused on activity recognition and analysis, while others have focused on fall detection. None of these works have focused on both of these challenges at the same time. Thus, it can be concluded that it is the need of the hour to leverage the immense potential at the intersection of ambient intelligence and the IoT to develop a framework that can not only track, study, analyze, and anticipate human behavior but also detect any anomalies, such as a fall or unconsciousness, that could constitute an emergency. It is also necessary that such systems are developed in way so that they are not environment-specific and can be seamlessly implemented and deployed in any IoT-based real world setting. This framework aimed to address these challenges by developing an approach for the analysis of human behavior at a fine-grain level with respect to the associated dynamic contextual, spatial, and temporal features to detect any anomalies that could constitute an emergency. The work involved the integration of advancements and technologies from multiple disciplines. This framework is introduced in Section 3, and a further discussion of how the salient features of this framework address these challenges and the drawbacks in the existing systems is presented in Section 5.

## 3. Proposed Work

In this section, we first present the steps towards the development of the functionality in our framework for the semantic analysis of user interactions on the context parameters during ADLs in order to identify a list of common behavioral patterns associated with different complex activities performed in any given IoT-based environment. In a real-world scenario, human activities are highly complex and involve multiple forms of user interactions that include a myriad of tasks and their dynamic characteristics, performed on the context parameters, based on the associated need related to the activity. Such complex real-world activities are referred to as complex activities. A complex activity can be broken down into atomic activities, context attributes, core atomic activities, core context attributes, other atomic activities, other context attributes, start atomic activities, end atomic activities, start context attributes, and end context attributes [30]. Here, atomic activities refer to the macro and micro level tasks and sub-tasks associated with the complex activity, and the environment parameters on which these atomic activities are performed are collectively known as context attributes. Those specific atomic activities that are crucial to a complex activity and without which the complex activity can never be completed are referred to as core atomic activities, and their associated context attributes are known as core context attributes. The atomic activities that are necessary to start a given complex activity are known as start atomic activities, and the atomic activities that are necessary to successfully end a given complex activity are known as end atomic activities. The context parameters on which these two types of atomic activities take place are known as start context attributes and end context attributes, respectively. All the atomic activities other than the core atomic activities are known as other atomic activities, and their associated context attributes are known as other context attributes. The semantic analysis of user interactions during complex activities involves analyzing all these characteristic features of activities with respect to contextual, spatial, and temporal information. The following are the steps for the development of this functionality in the proposed framework:

i.      Deploy both wireless and wearable sensors to develop an IoT-based interconnected environment.

ii.     Set up a data collection framework to collect the big data from these sensors during different ADLs performed in the confines of a given IoT-based space.

iii.    Use context-based user interaction data obtained from the wireless sensors to spatially map a given environment into distinct 'zones,' in terms of context attributes associated with distinct complex activities. Here, we define a 'zone' as a region in the user's spatial orientation where distinct complex activities take place. For instance, in the cooking zone, the complex activity of cooking could take place, but other complex activities like sleeping or taking a shower could not.

iv.     Analyze the atomic activities performed on different context attributes for a given complex activity, along with their characteristic features.

v.      Track user behavior in terms of joint point movements and joint point characteristics [31] for each atomic activity associated with any given complex activity.

vi.     Analyze the user behavior, atomic activities, and context attributes to form a general definition of a complex activity in each context-based spatial 'zone.'

vii.    Repeat (vi) for all the complex activities with respect to the context attributes as obtained from (iii) for a given IoT-based environment.

viii.   Analyze the activity definitions to find atomic activities and their characteristic features for all the complex activities associated with the different 'zones.'

ix.     Study the activity definitions to record the human behavior for all the atomic activities obtained from (viii).

x.      Analyze the behavior definitions in terms of joint point movements and characteristics to develop a knowledge base of common behaviors associated with all the complex activities in the different 'zones.'

xi.     Develop a dataset that consists of all these behavioral patterns and the big data from user interactions for each of these 'zones' in a given IoT-based environment.

xii.    Preprocess the data to detect and eliminate outliers and any noise prior to developing a machine learning model.

xiii.   Split the data into training and test sets and then test the machine learning model on the test set to evaluate its performance characteristics.

Upon the development of the above-discussed functionality in our framework, we implemented the following steps to develop the proposed intelligent decision-making algorithm that can detect emergencies or anomalies in user behavior based on studying the multimodal components of user interactions during complex activities in each 'zone.' Each 'zone' is associated with distinct complex activities that are further associated with a set of atomic activities, context attributes, core atomic activities, core context attributes, other atomic activities, other context attributes, start atomic activities, end atomic activities, start context attributes, and end context attributes. An analysis of the user behavior in terms of joint point characteristics [31] allows for the detection and analysis of these behavioral patterns and their relationships with the dynamic spatial features of the environment to detect any anomalies in user behavior that could constitute an emergency. For instance, the atomic activity of lying at night in the sleeping or bedroom zones could be interpreted as the person taking rest. However, the detection of the same atomic activity in the bathroom at the same time could indicate an emergency that could have resulted from a fall or unconsciousness. Such a situation would need the attention of caregivers or medical practitioners. The proposed intelligent decision-making algorithm was built on this concept for the detection of emergencies during complex activities, and the following are the steps for the development of this functionality:

i.　　Classify the complex activities from this dataset as per their relationships with atomic activities, context attributes, other atomic activities, other context attributes, core atomic activities, core context attributes, start atomic activities, end atomic activities, start context attributes, and end context attributes to develop semantic characteristics of complex activities.

ii.　　Track user movements to detect start atomic activities and start context attributes.

iii.　　If these detected start atomic activities and start context attributes match with the semantic characteristics of complex activities in the database, run the following algorithm: emergency detection from semantic characteristics of complex activities (EDSCCA).

iv.　　If these detected start atomic activities and start context attributes do not match with the semantic characteristics of complex activities in the knowledge base, then track the atomic activities, context attributes, other atomic activities, other context attributes, core atomic activities, core context attributes, start atomic activities, end atomic activities, start context attributes, and end context attributes to develop a semantic definition for a complex activity (SDCA).

v.　　If an SDCA is already present in the knowledge base, go to (vi), else update the database with the SDCA.

vi.　　Develop a dataset that consists of all these semantic definitions for complex activities and the big data from user interactions associated with them.

vii.　　Preprocess the data to detect and eliminate outliers and any noise prior to developing a machine learning model.

viii.　　Split the data into training and test sets and then test the machine learning model on the test set to evaluate its performance characteristics.

Next, we outline the steps for developing the proposed EDSCCA algorithm:

i.　　Track if the start atomic activity was performed on the start context attribute.

ii.　　Track if the end atomic activity was performed on the end context attribute.

iii.　　If (i) is true and (ii) is false:

　　a.　　Track all the atomic activities, context attributes, other atomic activities, other context attributes, core atomic activities, and core context attributes.

　　b.　　For any atomic activity or other atomic activity that does not match its associated context attribute, track the features of the user behavior.

　　c.　　If the user behavior features indicate lying and no other atomic activities are performed, the inference is an emergency.

iv.　　If (i) is true and (ii) is true:

　　a.　　The user successfully completed the activity without any emergency detected, so the inference is no emergency.

v.　　If (i) is false and (ii) is true:

　　a.　　Track all the atomic activities, context attributes, other atomic activities, other context attributes, core atomic activities, and core context attributes.

　　b.　　For any atomic activity or other atomic activity that does not match its associated context attribute, track the features of the user behavior.

　　c.　　If the user behavior features indicate lying and no other atomic activities performed, the inference is an emergency.

vi.　　If (i) is false and (ii) is false:

　　a.　　No features of human behavior were associated with the observed activities or, in other words, the user did not perform any activity, so the inference is no emergency.

We used one of our previous works [31] that presented a framework for human behavior representation in the context of ADLs based on joint point characteristics. These joint point characteristics primarily include joint point distances and joint point speeds. By studying these joint point characteristics associated with diverse behavioral patterns, this framework tracks the dynamic changes in the skeleton that point to interpretations of human pose and posture. The dynamics of human pose and posture are then used by the framework to analyze human behavior and its associated features during multimodal interactions in the context of ADLs. This concept is outlined in Figure 2. According to this methodology, each point on the skeletal tracking, as obtained from a Microsoft Kinect sensor, is assigned a joint number and a definition based on the kind of underlining movements associated with that joint point. The associated joint point characteristics, in terms of the individual joint point speeds and the distance between two or more joint points, undergo changes based on the behavioral patterns of the user. We applied this concept to analyze the ADLs in terms of the atomic activities, context attributes, other atomic activities, other context attributes, core atomic activities, and core context attributes in order to identify the list of behavioral patterns associated with each of these ADLs. This analysis also involved modelling all possible instances of each complex activity while assigning weights to the individual atomic activities, context attributes, other atomic activities, other context attributes, core atomic activities, and core context attributes based on probabilistic reasoning. This was done by using Equations (1)–(3), which were proposed in [32].

$$\alpha = a_t C0 + a_t C1 + a_t C2 + \ldots \ldots a_t Ca_t = 2^{a_t} \tag{1}$$

$$\beta = (a_t - c_t)C0 + (a_t - ct)C1 + (a_t - c_t)C2 + \ldots + (a_t - c_t)C(a_t - c_t) = 2^{(at-ct)} \tag{2}$$

$$\gamma = 2at - 2^{(at-ct)} = 2^{(at-ct)} * (2^{ct} - 1) \tag{3}$$

where $\alpha$ represents all possible ways by which any complex activity can be performed including false starts; $\beta$ represents all the ways of performing any complex activity where the user always reaches the end goal; $\gamma$ represents all the ways of performing any complex activity where the user never reaches the end goal; Ati represents all the atomic activities related to the complex activity, where i is a positive integer; Cti represents all the context attributes related to the complex activity, where i is a positive integer; AtS represents a list of all the Ati that are start atomic activities; CtS represents a list of all the Cti that are start context attributes; AtE represents a list of all the Ati that are end atomic activities; CtE represents a list of all the Cti that are end context attributes; $\gamma$At represents a list of all the Ati that are core atomic activities; $\rho$Ct represents a list of all the Cti that are core context attributes; $a_t$ represents the number of Ati related to the complex activity; $b_t$ represents the number of Cti related to the complex activity; $c_t$ represents the number of $\gamma$At related to the complex activity; and $d_t$ represents the number of $\rho$Ct related to the complex activity.

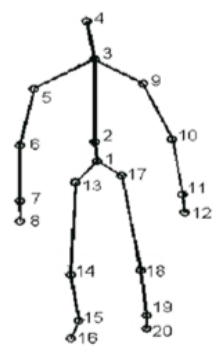

| Joint | Definition | Joint | Definition |
|-------|------------|-------|------------|
| 1 | Center of Hip | 11 | Right Wrist |
| 2 | Spine | 12 | Right Hand |
| 3 | Center of Shoulder | 13 | Left Hip |
| 4 | Head | 14 | Left Knee |
| 5 | Left Shoulder | 15 | Left Ankle |
| 6 | Left Elbow | 16 | Left Foot |
| 7 | Left Wrist | 17 | Right Hip |
| 8 | Left Hand | 18 | Right Knee |
| 9 | Right Shoulder | 19 | Right Ankle |
| 10 | Right Elbow | 20 | Right Foot |

**Figure 2.** The methodology to represent skeletal tracking in terms of joint points and their associated definitions [31].

The work in [32] presented a mathematical foundation for modelling all possible user interactions related to atomic activities, context attributes, other atomic activities, other context attributes, core atomic activities, and core context attributes associated with any given complex activity. The objective of the work in [32] was to develop a knowledge base that would consist of all possible tasks and actions performed on context parameters, related to any given complex activity, arising from universal diversity and the variations in the context parameters based on the associated spatial and temporal characteristics of user interactions. In this work, these equations were developed by integrating complex activity analysis [30], the principles of the binomial theorem [33], and permutation and combination principles. These equations represent the diverse ways by which a complex activity may be performed. Equation (1) represents all possible ways by which a complex activity can be modelled, including distractions, false starts, one or more missed Ati, one or more missed Cti, one or more missed AtS, one or more missed CtS, one or more missed AtE, one or more missed CtE, one or more missed $\gamma$At, and one or more missed $\rho$Ct. Equation (2) represents all those scenarios where the user reached the end goal or, in other words, the user performed all the $\gamma$At on the $\rho$Ct related to a given complex activity. Equation (3) represents all those scenarios where the user did not perform one or more $\gamma$At on the $\rho$Ct related to a given complex activity, as well as one or more missed AtS, one or more missed CtS, one or more missed AtE, and one or more missed CtE. Weights were assigned to the individual Ati and Cti by probabilistic reasoning principles, as outlined in [30]. The weights indicate the relevance or importance of the task or action towards helping the user reach the end goal or desired outcome. A higher value of the weight indicates a greater relevance, and a lower value of the weight indicates a lesser relevance of the associated Ati and Cti. The $\gamma$At and $\rho$Ct are assigned the highest weights as compared to the other Ati and Cti. The weights associated with all the Ati and Cti can be analyzed to determine the threshold weight of the complex activity, which determines whether a given complex activity was properly performed. Here, properly performed refers to whether the user was able to successfully reach the end goal or outcome associated with a given complex activity. The threshold weight varies based on the nature and number of AtS, CtS, AtE, CtE, $\gamma$At, and $\rho$Ct related to a complex activity. Each instance of a complex activity, denoted by Equation (1), is also assigned a different weight based on the number of AtS, CtS, AtE, CtE, $\gamma$At, and $\rho$Ct, as well as the nature and sequence in which these actions were performed. When this weight exceeds the threshold weight, it indicates that the user reached the end goal, and such activity instances are represented by Equation (2). Table 1 outlines the analysis for a typical ADL, eating lunch, as described by this methodology. In Table 2, we represent the analysis of this complex activity as per Equations (1)–(3) to study the characteristics of the associated Ati, Cti, AtS, CtS, AtE, CtE, $\gamma$At, $\rho$Ct, $a_t$, $b_t$, $c_t$, and $d_t$.

**Table 1.** Analysis of the complex activity of eating lunch in terms of joint point characteristics [31].

| Atomic Activities | Context Attributes | Joint Points Pairs That Experience Change |
|---|---|---|
| **At1**: Standing (0.08) | **Ct1**: Lights on (0.08) | No change |
| **At2**: Walking towards dining table (0.20) | **Ct2**: Dining area (0.20) | (13,17), (14,18), (15,19), and (16,20) |
| **At3**: Serving food on a plate (0.25) | **Ct3**: Food present (0.25) | (7, 11) and (8,12) |
| **At4**: Washing hand/using hand sanitizer (0.20) | **Ct4**: Plate present (0.20) | (7, 11) and (8,12) |
| **At5**: Sitting down (0.08) | **Ct5**: Sitting options available (0.08) | No change |
| **At6**: Starting to eat (0.19) | **Ct6**: Food quality and taste (0.19) | (6,3), (7,3), (8,3), (6,4), (7,4), (8,4) or (10,3), (11,3), (12,3), (10,4), (11,4), and (12,4) |

As can be seen from Table 2, where $\alpha$ = 64, there are 64 different ways by which this complex activity can be performed. However, the value of $\gamma$ = 60 means that 60 out of these 64 ways would not lead to the end goal or the desired outcome. The remaining activity instances indicated by $\beta$ = 4 refers to those instances when the user would always reach the end goal of this complex activity. One such instance is shown in Table 1.

**Table 2.** Analyzing multiple characteristics of a typical complex activity—eating lunch.

| Complex Activity Characteristics | Value(s) |
|---|---|
| Ati, all the atomic activities related to the complex activity | At1, At2, At3, At4, At5, and At6 |
| Cti, all the context attributes related to the complex activity | Ct1, Ct2, Ct3, Ct4, Ct5, and Ct6 |
| AtS, list of all the Ati that are start atomic activities | At1 and At2 |
| CtS, list of all the Cti that are start context attributes | Ct1 and Ct2 |
| AtE, list of all the Ati that are end atomic activities | At5 and At6 |
| CtE, list of all the Cti that are end context attributes | Ct5 and Ct6 |
| $\gamma$At, list of all the Ati that are core atomic activities | At2, At3, At4, and At6 |
| $\rho$Ct, list of all the Cti that are core context attributes | Ct2, Ct3, Ct4, and Ct6 |
| $a_t$, number of Ati related to the complex activity | 6 |
| $b_t$, number of Cti related to the complex activity | 6 |
| $c_t$, number of $\gamma$At related to the complex activity | 4 |
| $d_t$, number of $\rho$Ct related to the complex activity | 4 |
| $\alpha$, all possible ways by which any complex activity can be performed including false starts | 64 |
| $\beta$, all the ways of performing any complex activity where the user always reaches the end goal | 4 |
| $\gamma$, all the ways of performing any complex activity where the user never reaches the end goal | 60 |

To develop this framework, we used an open-source dataset [34] that contains the big data of user interactions recorded during multiple ADLs in an IoT-based environment. The complex activities and their associated characteristics in this dataset can be distinctly mapped to four spatial 'zones'—kitchen, bedroom, office, and toilet—in the simulated and interconnected IoT-based environment. The big data in this dataset consisted of data attributes that provided the location, or the 'zone'-related data associated with all these ADLs. These data were used to analyze the indoor location of the user with respect to the context attributes of interest for a given complex activity in the IoT-based environment. The context attributes associated with different instances of each of these ADLs were studied by the approach discussed in Tables 1 and 2. The dataset also consisted of accelerometer and gyroscope data that were collected from wearables and that represented diverse behavioral patterns during different instances of each of the ADLs performed in each of these spatial 'zones.' These data were used to study, analyze, and interpret the multimodal characteristics of human behavior distinct to different complex activities. Here, as per the data and their characteristics present in the dataset, we defined lying and being unable to get up in any other location other than a bedroom as an emergency. This definition of an emergency can also be modified, e.g., to detect a long lie, as per the complex activities and their semantic characteristics for a given IoT-based environment.

We used RapidMiner, previously known as Yet Another Learning Environment (YALE) [35], for the development of this framework. RapidMiner is a software tool that consists of several built-in functions known as 'operators' that can be used to implement a range of computational functions including machine learning, artificial intelligence, and natural language processing algorithms. The tool also allows for the seamless customization of these 'operators' as per the needs of the model being developed. Multiple 'operators' can be put together in the tool to develop an application, which is known as a 'process.' There are two versions of RapidMiner available—the free version and the paid version. The free version has a processing limit of 10,000 rows. The dataset that we used for this study did not exceed 10,000 rows, so this limitation of the free version did not affect our results and findings. The version of RapidMiner that we used was 9.8.000, and the same was run on a Microsoft Windows 10 computer with an Intel (R) Core(TM) i7-7600U CPU @ 2.80 GHz, 2 core(s) and 4 logical processor(s) for the development and implementation of the proposed framework.

## 4. Results

In this section, we present the results obtained from the proposed framework by using the dataset [34]. The big data present in the dataset represented various kinds of ADLs—sleeping, changing clothes, relaxing, cooking, eating, working, and defecating, as well as emergency situations in the kitchen, bedroom, office, and toilet. The emergency corresponded to the user lying on the ground in any location other than the bedroom, which could have resulted from a fall or unconsciousness. As per the methodology discussed in Figure 2 and Tables 1 and 2, we developed definitions of all the complex activities that occurred in a given IoT-space. Then, we developed a process in RapidMiner to identify and interpret the list of common behavioral patterns associated with each of these ADLs in this dataset, performed in the spatial locations or 'zones'—bedroom, kitchen, office, and toilet. We used the 'Dataset' operator to import this dataset into RapidMiner. The 'Data Preprocessing' operator was used to preprocess the data and to study the various characteristics of human behavior as outlined in Section 3. The data processing involved the studying, analysis, and interpretation of the dynamic characteristics of human behavior data associated with the diverse complex activities performed in each of the spatial 'zones' represented in the dataset. The dataset that we used for these pre-processing steps consisted of 295 rows. First, we studied the different ADLs performed in each of these 'zones'—bedroom, kitchen, office, and toilet. This is shown in Figure 3, where the location or 'zone' is plotted on the *x*-axis, and the different ADLs are represented on the *y*-axis. As there were nine different ADLs, so we represented each ADL with a different color; this color coding is mentioned in the figure. Each occurrence of an ADL in a specific 'zone' is represented with a bubble corresponding to that zone. For instance, in the toilet zone, the activities of defecating and emergency were observed, so these two activities were tracked using distinct colors for this 'zone.'

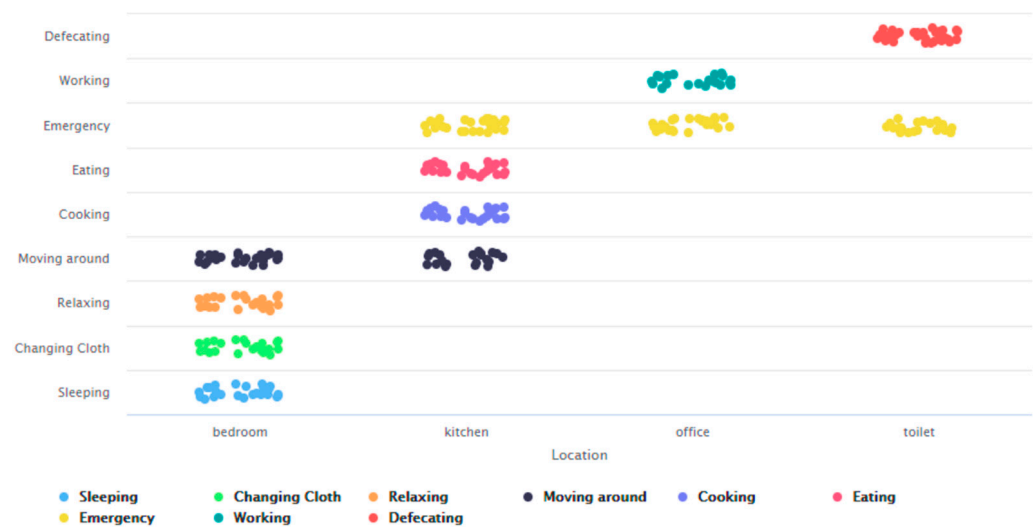

**Figure 3.** Analysis of different ADLs performed in the different spatial locations or 'zones' in a given Internet of Things (IoT)-based environment.

After detecting and studying the different ADLs local to each 'zone,' we studied the associated atomic activities, context attributes, other atomic activities, other context attributes, core atomic activities, and core context attributes associated with each of these ADLs to study the common behavioral patterns local to each ADL in each zone that we observed from Figure 3. This analysis is shown in Figure 4, where the *x*-axis represents the location, and the common behavioral patterns of lying, standing, sitting, and walking are represented on the *y*-axis. As there were multiple ADLs to which these common behavioral patterns belonged, so we represented each ADL by using a different color. Each occurrence of an ADL in a specific 'zone' is represented with a bubble corresponding to that zone. For instance, from Figure 3, we can observe that in the toilet zone, the activities of defecating and emergency occur multiple times. The behavioral patterns associated with these activities are sitting and lying, so these behaviors were represented using different colors, as shown in Figure 4.

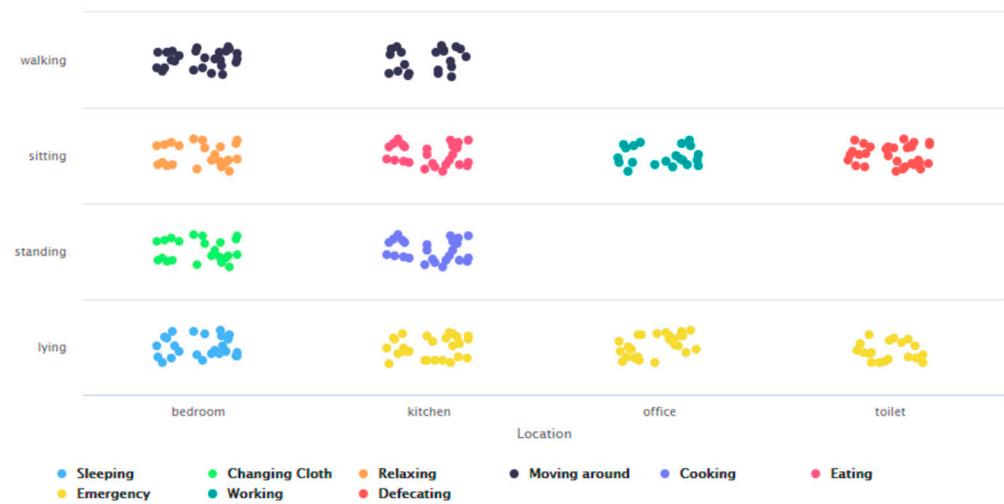

**Figure 4.** Representation of common and distinct behavioral patterns associated with the different ADLs performed in the different spatial locations or 'zones' in a given IoT-based environment.

After studying these activity patterns distinct to different ADLs local to each zone, we studied the characteristics of the human behaviors at a fine-grain level associated with each of these ADLs. This was done by analyzing the accelerometer and gyroscope data corresponding to occurrences of each of the common behavioral patterns—lying, standing, sitting, and walking—for different ADLs in each of these spatial 'zones.' The study and analysis of the accelerometer and gyroscope data for these common behavioral patterns for all these ADLs performed in the kitchen, bedroom, office area, and toilet are shown in Figures 5–8, respectively. In each of these figures, the common behavioral patterns are plotted on the *x*-axis. The *y*-axis represents the accelerometer data and gyroscope in the *x*, *y*, and *z* directions, each of which is plotted with a distinct color.

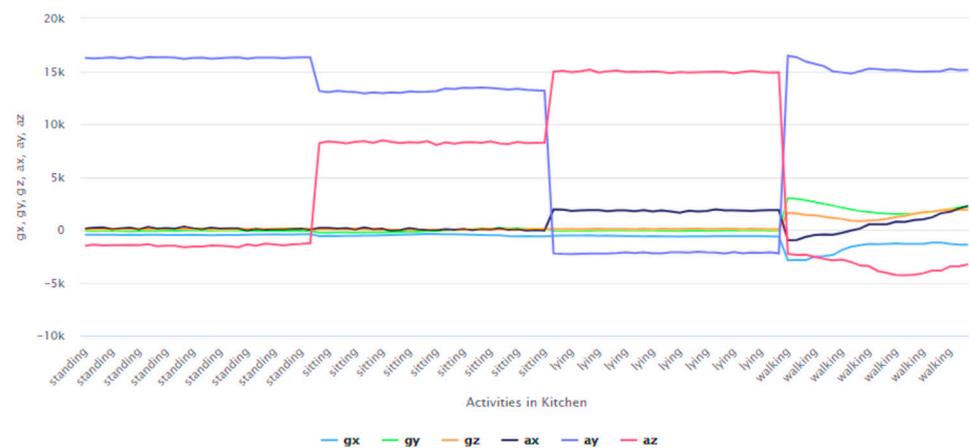

**Figure 5.** The study and analysis of the accelerometer and gyroscope data for the common behavioral patterns—lying, standing, sitting, and walking—for all ADLs performed in the kitchen. Due to paucity of space, analyses of a some of the ADLs are shown here.

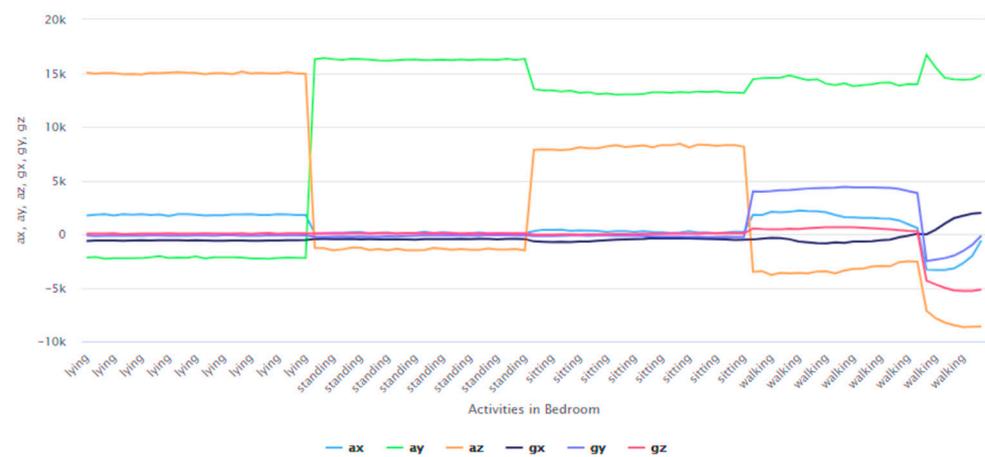

**Figure 6.** The study and analysis of the accelerometer and gyroscope data for the common behavioral patterns—lying, standing, sitting, and walking—for all ADLs performed in the bedroom. Due to paucity of space, analyses of a some of the ADLs are shown here.

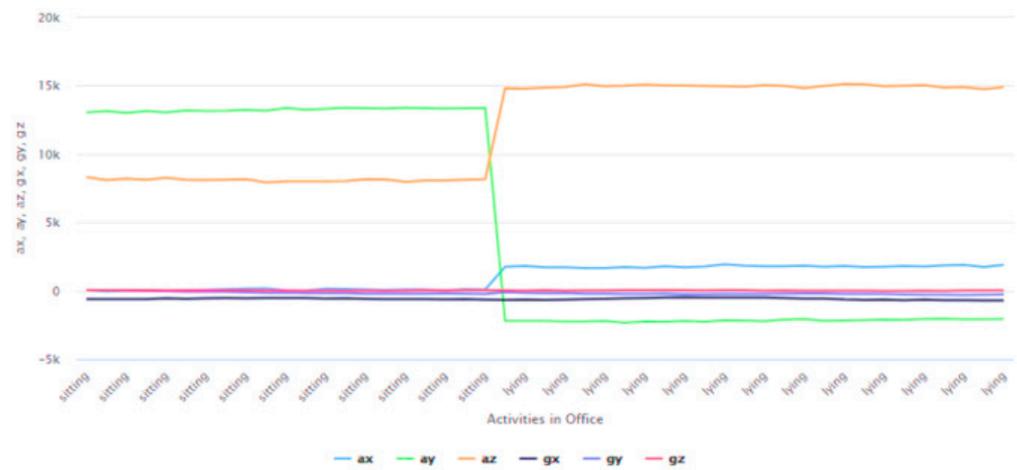

**Figure 7.** The study and analysis of the accelerometer and gyroscope data for the common behavioral patterns—lying, standing, sitting, and walking—for all ADLs performed in the office area. Due to paucity of space, analyses of a some of the ADLs are shown here.

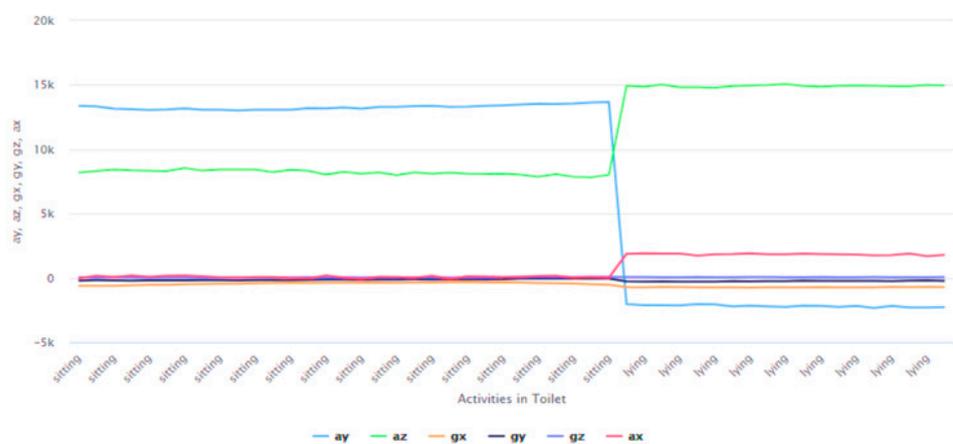

**Figure 8.** The study and analysis of the accelerometer and gyroscope data for the common behavioral patterns—lying, standing, sitting, and walking—for all ADLs performed in the toilet. Due to paucity of space, analyses of a some of the ADLs are shown here.

After performing this analysis, we used the 'Split Data' operator to split the data into training and test sets; 75% of the data were used for training, and the remaining 25% were used for testing. We used a k-NN classifier to develop the machine-learning functionality of our framework. k-NN [36] is a non-parametric machine learning classifier. k-NN works by comparing an unknown data sample to 'k' closest training examples in the dataset to classify the unknown data into one of these samples. Here, closeness refers to the distance in a space represented by 'p,' where 'p' is the number of attributes in the training set. There are various approaches for the calculation of this distance. For the development of the proposed approach, we used the Euclidean distance approach in RapidMiner [35]. The Euclidean distance [37] between two points 'm' and 'n' is computed as shown in Equation (4):

$$d(m,n) = \sqrt{\sum_{i=1}^{p}(m_i - n_i)^2} \qquad (4)$$

where m and n are two points in the Euclidean space, d (m,n) represents the distance between the two points m and n in the Euclidean space, $m_i$ represents the vector in the Euclidean space that connects the point m to the origin, $n_i$ represents the vector in the Euclidean space that connects the point n to the origin, and p represents the p-space.

The k-NN model that we developed consisted of 11 nearest neighbors. The model was developed using 222 examples consisting of three dimensions of each of the activity classes representing lying, standing, walking, and sitting. We tested the classifier by using the 'Apply Model' operator and evaluated its performance characteristics by using the 'Performance' operator. This RapidMiner process is shown in Figure 9, and the order in which the 'operators' associated with this RapidMiner process were executed when the process was compiled and run is shown in Figure 10. Thereafter, we studied the effectiveness and performance characteristics of our framework to detect these behavioral patterns—walking, sleeping, sitting, and lying—in different spatial locations. The RapidMiner process studied each row of the test dataset, which constituted a user interaction with the context parameters and detected the associated behavioral patterns.

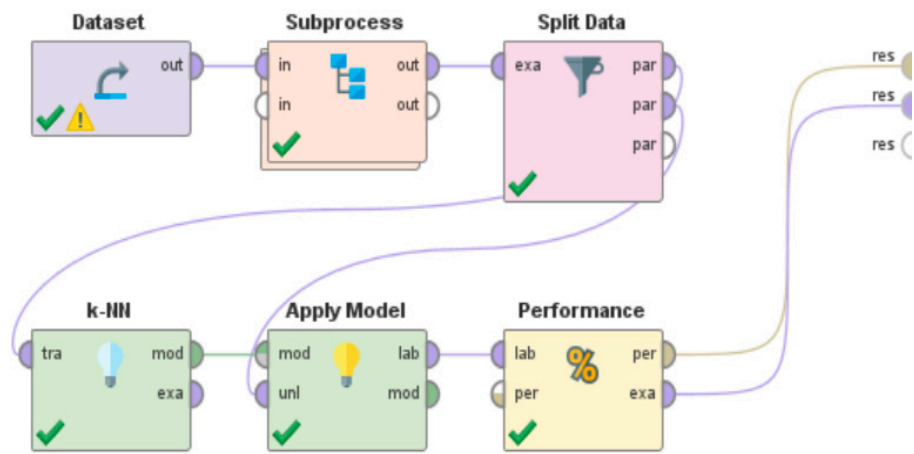

**Figure 9.** RapidMiner process for the development of the functionality of the framework to perform the semantic analysis of user interactions on the diverse context parameters during ADLs to identify a list of distinct behavioral patterns associated with different complex activities performed in a given IoT-based environment.

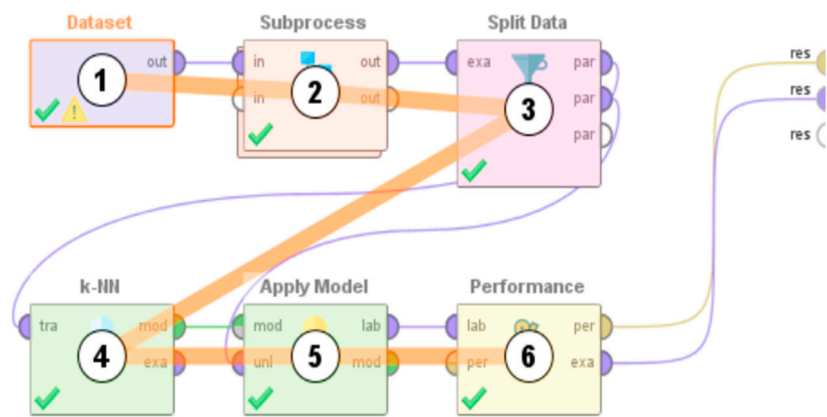

**Figure 10.** The order of execution of all the operators upon the compilation and execution of the RapidMiner process shown in Figure 9.

The output of this RapidMiner process was in the form of a table where each row consisted of the attributes as outlined in Table 3. Here, the degree of certainty expresses the certainty of prediction of the associated behavioral pattern of the user by the developed k-NN-based machine learning model. To predict the same, the k-NN model in RapidMiner assigned a confidence value to each of these behavioral patterns, and the behavior with the highest confidence was the final prediction of the model for that specific user interaction. For instance, in row number 2, the confidence values associated with lying, standing, sitting, and walking are 0.818, 0.182, 0, and 0, respectively, so the prediction of the model was lying. This output table had 73 rows, but only the first 13 rows are shown in Figure 11.

**Table 3.** Description of the attributes of the output of the RapidMiner process shown in Figure 11.

| Attribute Name | Description |
| --- | --- |
| Row No | The row number in the output table |
| Activity | The actual behavioral pattern associated with a given ADL |
| Prediction (Activity) | The predicted behavioral pattern associated with a given ADL |
| Confidence (lying) | The degree of certainty that the user was lying during this ADL |
| Confidence (standing) | The degree of certainty that the user was standing during this ADL |
| Confidence (sitting) | The degree of certainty that the user was sitting during this ADL |
| Confidence (walking) | The degree of certainty that the user was sitting during this ADL |

| Row No. | Activity | prediction(Activity) | confidence(lying) | confidence(standing) | confidence(sitting) | confidence(walking) |
| --- | --- | --- | --- | --- | --- | --- |
| 1 | lying | lying | 0.723 | 0.277 | 0 | 0 |
| 2 | lying | lying | 0.818 | 0.182 | 0 | 0 |
| 3 | lying | lying | 0.645 | 0.355 | 0 | 0 |
| 4 | standing | sitting | 0.089 | 0.272 | 0.639 | 0 |
| 5 | standing | standing | 0.447 | 0.553 | 0 | 0 |
| 6 | standing | lying | 0.637 | 0.363 | 0 | 0 |
| 7 | standing | lying | 0.630 | 0.370 | 0 | 0 |
| 8 | sitting | sitting | 0 | 0 | 1 | 0 |
| 9 | sitting | sitting | 0 | 0 | 1.000 | 0 |
| 10 | sitting | sitting | 0.088 | 0 | 0.912 | 0 |
| 11 | sitting | sitting | 0.172 | 0 | 0.828 | 0 |
| 12 | sitting | sitting | 0 | 0 | 1 | 0 |
| 13 | sitting | sitting | 0.087 | 0 | 0.913 | 0 |

**Figure 11.** The output table of the RapidMiner process shown in Figure 3 for the detection of distinct behavioral patterns associated with the different ADLs. This output table had 73 rows, but only the first 13 rows are shown here.

The performance accuracy of this model was evaluated by using a confusion matrix, where both the overall performance and the individual class precision values were computed. Figures 12 and 13 show the tabular representation and plot view of the confusion matrix, respectively. A confusion matrix [38] is a method of evaluating and studying the performance characteristics of a machine learning-based algorithm. The number of instances of a data label in the predicted class is represented by each row of the matrix, and the number of instances of a data label in the actual class is represented by each column of the matrix. The matrix can also be inverted to have the rows represent the columns and vice versa. Such a matrix allows for the calculation of multiple performance characteristics associated with the machine learning model. These include overall accuracy, individual class precision values, recall, specificity, positive predictive values, negative predictive values, false positive rates, false negative rates, and F-1 scores. To evaluate the performance characteristics of our proposed approach, we focused on two of these performance metrics—the overall accuracy and the individual class precision values, which are calculated by the formula as shown in Equations (5) and (6), respectively:

$$\text{Acc} = \frac{\text{True(P)} + \text{True(N)}}{\text{True(P)} + \text{True(N)} + \text{False(P)} + \text{False(N)}} \tag{5}$$

$$\text{Pr} = \frac{\text{True(P)}}{\text{True(P)} + \text{False(P)}} \tag{6}$$

where Acc is the overall accuracy of the machine-learning model, Pr is the class precision value, True(P) means true positive, True(N) means true negative, False(P) means false positive, and False(N) means false negative.

**accuracy: 76.71%**

| | true lying | true standing | true sitting | true walking | class precision |
|---|---|---|---|---|---|
| pred. lying | 19 | 8 | 3 | 0 | 63.33% |
| pred. standing | 1 | 3 | 0 | 0 | 75.00% |
| pred. sitting | 4 | 1 | 22 | 0 | 81.48% |
| pred. walking | 0 | 0 | 0 | 12 | 100.00% |
| class recall | 79.17% | 25.00% | 88.00% | 100.00% | |

**Figure 12.** The performance accuracy (studied via a confusion matrix—tabular view) of the Rapid-Miner process shown in Figure 9 for the detection of distinct behavioral patterns associated with the different ADLs performed in the different spatial locations in a given IoT-based environment.

As can be seen from Figures 12 and 13, this machine learning model achieved an overall performance accuracy of 76.71%, with respective class precision values for lying, standing, sitting and walking of 63.33%, 75.00%, 81.48%, and 100.00%. Our understanding is that out of lying, standing, sitting and walking, only walking constitutes a movement of the user from one location to the other, which is distinct compared to the other behaviors on the dataset—lying, sitting, and standing. This makes the associated user interactions and behavior-related data very different from the other behaviors. Thus, the detection of walking by the machine learning model could achieve 100.00% accuracy.

Thereafter, we developed the other functionality of our framework—the intelligent decision-making algorithm that can analyze these behavioral patterns and their relationships with the dynamic spatial features of the environment to detect any anomalies in user behavior that could constitute an emergency, as outlined in Section 2. This functionality of our framework was developed as a RapidMiner 'process' that is shown in Figure 14, and the order in which the various operators of this RapidMiner process were executed upon the compilation of the same is shown in Figure 15. For the purpose of evaluating the efficacy of this framework, we were interested in developing a binary classifier that could classify a situation as an 'emergency' or 'non-emergency.' Thus, all instances of activities other than an emergency were labelled as 'non-emergency' in this dataset for the development of this RapidMiner 'process.' The 'Dataset' operator allowed for the importation of the data into the RapidMiner platform for developing this 'process.' The 'Set Role' operator was used to inform RapidMiner of the data attribute and its characteristics that should be predicted. In this case, it was either 'emergency' or 'non-emergency.' The 'Data Processing' operator was used to implement the knowledge base and make the model aware of the rest of the relationships and dependencies amongst the data attributes as per the characteristics of our framework and the proposed EDSCCA. The 'Data Preprocessing' operator also consisted of the of the 'Split Data' operator, which was used to split the data into training and test sets. We used 75% of the data for training and 25% of the data for testing after the removal of the outliers, as per the data preprocessing steps outlined in Section 3. Next, we used a k-NN classifier to develop this binary classification model. This k-NN classifier was also developed based on the Euclidean distance approach represented in Equation (4). This classification model consisted of five nearest neighbors and 186 examples with eight dimensions of the two classes—emergency and non-emergency. The 'Apply Model' operator was used to apply this learning model to the test data. The 'Performance' operator was used to evaluate the performance characteristics of the learning model. For the performance metrics, we used the confusion matrix to study the overall accuracy of the model, as well as the individual class precision values.

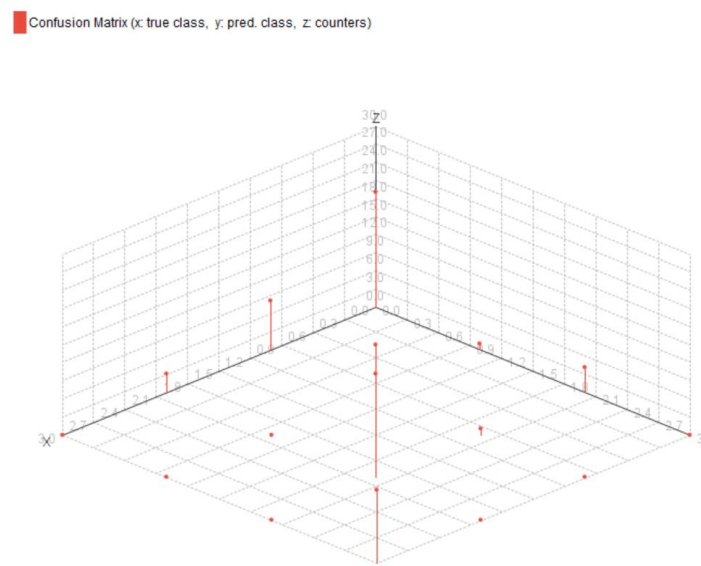

**Figure 13.** The performance accuracy (studied via a confusion matrix—plot view) of the RapidMiner process shown in Figure 9 for the detection of distinct behavioral patterns associated with different ADLs performed in the different spatial locations in a given IoT-based environment.

This RapidMiner process studied each row of the dataset, which consisted of different behavioral patterns associated with an ADL, to classify the associated behavior as emergency or non-emergency.

The output of this RapidMiner process was in the form of a table where each row consisted of the attributes outlined in Table 4. Here, the degree of certainty expresses the certainty of prediction of emergency or non-emergency by the developed k-NN-based machine learning model. To predict the same, the k-NN classification model in RapidMiner assigned a confidence value to each of these behavioral patterns, and the behavior with the highest confidence value was the final prediction of the model for that specific user interaction. For instance, in row number 2, the confidence values associated with non-emergency and emergency are 0.811 and 0.189, respectively, so the prediction of the model was non-emergency for the specific user interaction represented by this row. This output table had 62 rows, but only the first 13 rows are shown in Figure 16.

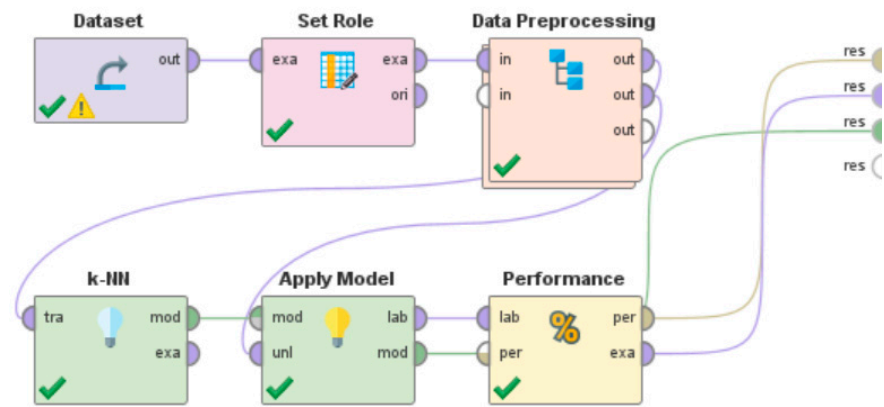

**Figure 14.** RapidMiner process for the development of the intelligent decision-making algorithm of the framework that can analyze distinct behavioral patterns and their relationships with the dynamic contextual and spatial features of the environment to detect any anomalies in user behavior that could constitute an emergency.

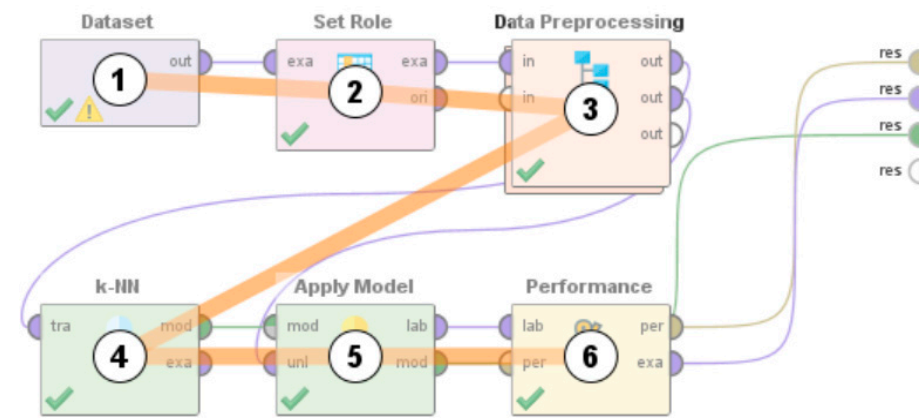

**Figure 15.** The order of execution of all the operators upon the compilation and execution of the RapidMiner process shown in Figure 14.

**Table 4.** Description of the attributes of the output of the RapidMiner process shown in Figure 16.

| Attribute Name | Description |
| --- | --- |
| Row No | The row number in the output table |
| Complex Activity | The actual user behavior (either emergency or non-emergency) associated with a given complex activity (ADL) |
| Prediction (Complex Activity) | The predicted user behavior (either emergency or non-emergency) associated with a given complex activity (ADL) |
| Confidence (Non-Emergency) | The degree of certainty that the user behavior associated with a given complex activity did not constitute an emergency |
| Confidence (Emergency) | The degree of certainty that the user behavior associated with a given complex activity constituted an emergency |

| Row No. | Complex Activity | prediction(Complex Activity) | confidence(Non-Emergency) | confidence(Emergency) |
|---------|------------------|------------------------------|---------------------------|------------------------|
| 1 | Non-Emergency | Non-Emergency | 0.801 | 0.199 |
| 2 | Non-Emergency | Non-Emergency | 0.811 | 0.189 |
| 3 | Non-Emergency | Non-Emergency | 0.616 | 0.384 |
| 4 | Non-Emergency | Emergency | 0.220 | 0.780 |
| 5 | Non-Emergency | Emergency | 0.403 | 0.597 |
| 6 | Non-Emergency | Non-Emergency | 0.815 | 0.185 |
| 7 | Non-Emergency | Non-Emergency | 1 | 0 |
| 8 | Non-Emergency | Emergency | 0.393 | 0.607 |
| 9 | Non-Emergency | Non-Emergency | 1 | 0 |
| 10 | Non-Emergency | Non-Emergency | 1 | 0 |
| 11 | Non-Emergency | Non-Emergency | 1 | 0 |
| 12 | Non-Emergency | Non-Emergency | 1 | 0 |
| 13 | Non-Emergency | Non-Emergency | 1 | 0 |

**Figure 16.** The output table of the intelligent decision-making algorithm of the framework, developed as a RapidMiner process, that can analyze distinct behavioral patterns and their relationships with the dynamic contextual and spatial features of the environment to detect any anomalies in user behavior that could constitute an emergency. This output table had 62 rows, but only the first 13 rows are shown here.

The performance characteristics of this framework were evaluated in the form a confusion matrix, as shown in Figures 17 and 18, with Figure 17 representing the tabular view and Figure 18 representing the plot view of the confusion matrix. By using the confusion matrix, both the overall performance and individual class precision performance values were computed.

As can be observed from Figures 17 and 18, the framework achieved an overall performance accuracy of 83.87%, with the sub-class precision for the detection of 'non-emergency' being 85.42% and the sub-class precision for the detection of 'emergency' being 78.57%.

**accuracy: 83.87%**

| | true Non-Emergency | true Emergency | class precision |
|---|---|---|---|
| pred. Non-Emergency | 41 | 7 | 85.42% |
| pred. Emergency | 3 | 11 | 78.57% |
| class recall | 93.18% | 61.11% | |

**Figure 17.** The performance accuracy (studied via a confusion matrix—tabular view) of the Rapid-Miner process shown in Figure 14 that involves the development of the intelligent decision-making algorithm of the framework that can analyze distinct behavioral patterns and their relationships with the dynamic contextual and spatial features of the environment to detect any anomalies in user behavior that could constitute an emergency.

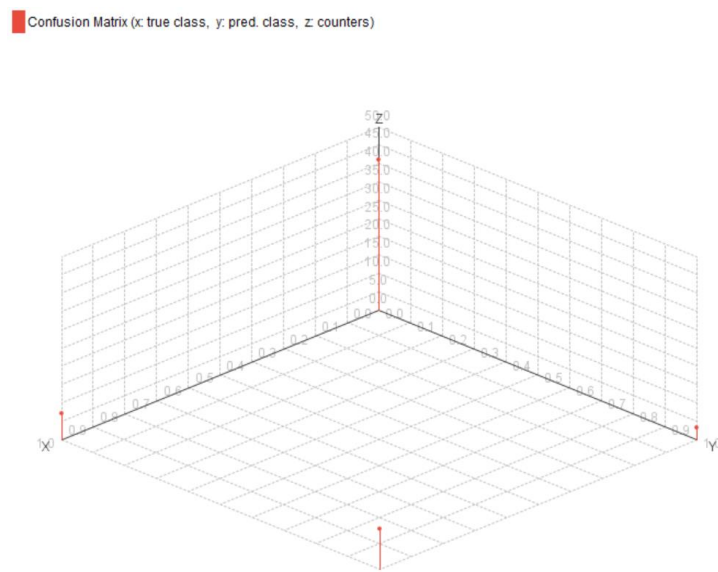

**Figure 18.** The performance accuracy (studied via a confusion matrix—plot view) of the RapidMiner process shown in Figure 14 that involves the development of the intelligent decision-making algorithm of the framework that can analyze distinct behavioral patterns and their relationships with the dynamic contextual and spatial features of the environment to detect any anomalies in user behavior that could constitute an emergency.

## 5. Comparative Discussion

Despite several advances and emerging technologies in the fields of human activity recognition, human behavior analysis, and their related application domains, the existing systems [9–29] have several limitations and drawbacks, as outlined in Section 2. This framework, which integrates the latest advancements and technologies in human–computer interaction, machine learning, Internet of Things, pattern recognition, and ubiquitous computing, aims to take a rather comprehensive approach to addressing these challenges in this field. In this section, we discuss these specific challenges and outline how our framework addresses the same and outperforms these existing systems in terms of their technical characteristics, functionalities, and operational features. This is presented as follows:

1.  Several researchers in this field have only focused on activity recognition and that too at a superficial level. Various methodologies such as sensor technology-driven [9], RGB frame-based [10], hidden Markov model-based [11], and computer vision-based [12] methodologies have been proposed by researchers, but the main limitation of such systems is their inability to analyze complex activities at a fine-grain level to interpret the associated dynamics of user interactions and their characteristic features. Our framework addresses this challenge by being able to perform the semantic analysis of user interactions with diverse contextual parameters during ADLs. By semantic analysis, we refer to the functionalities of our framework to (1) analyze complex activities in terms of the associated postures and gestures, which are interpreted in terms of the skeletal joint point characteristics (Figure 1); (2) interpret the interrelated and interdependent relationships between atomic activities, context attributes, core atomic activities, core context attributes, other atomic activities, other context attributes, start atomic activities, end atomic activities, start context attributes, and end context attributes associated with any complex activity (Table 1); (3) detect all possible dynamics of user interactions and user behavior that could be associated with any complex activity (Table 2); (4) identify a list of distinct fine-grain level behavioral patterns—walking, sleeping, sitting, and lying—associated with different complex activities (Figures 9 and 11), which achieved a performance accuracy of 76.71% when tested on a dataset of ADLs (Figures 12 and 13); and (5) use an intelli-gent decision-making algorithm that can analyze these distinct behavioral patterns and their relationships with the dynamic contextual and spatial features of the environment to detect any anomalies in user behavior that could constitute an emergency (Figures 14 and 16), which achieved an overall performance accuracy of 83.87% when tested on a dataset of ADLs (Figures 17 and 18).

2. Some of the recent works that have focused on activity analysis were limited to certain tasks and could not be generalized for different activities. For instance, in [17], the work focused on eating activity recognition and analysis; in [13], the activity analysis was done to detect enter and exit motions only in a given IoT-based space. In [18], the methodology focused on the detection of simple and less complicated activities, such an cooking, and [22] presented a system that could remind its users to take their routine medications. The analysis of such small tasks and actions are important, but the challenge in this context is the fact that these systems are specific to such tasks and cannot be deployed or implemented in the context of other activities. With its functionalities to perform complex activity recognition and analysis of skeletal joint point characteristics, our framework can analyze and interpret any complex activity and its associated tasks and actions, thereby addressing this challenge. When tested on a dataset, our framework was able to recognize and analyze all nine complex activities—sleeping, changing cloth, relaxing, moving around, cooking, eating, emergency, working, and defecating—that were associated with this dataset. It is worth mentioning here that our framework cannot only recognize these specific nine complex activities, because its characteristics allow it to recognize and analyze any set of complex activities represented by the big data associated with user interactions in a given IoT-based context, which could be a from a dataset or from a real-time sensor-based implementation of the IoT framework.

3. A number of these methodologies have focused on activities in specific settings and cannot be seamlessly deployed in other settings consisting of different context parameters and environment variables. For instance, in [14,16], the presented systems are specific to hospital environments, the methodology presented in [21] is only applicable to a kitchen environment, and the approach in [28] is only applicable to a workplace environment. While such systems are important for safe and assisted living experiences in these local spatial contexts, their main drawback is the fact that these tools are dependent on the specific environmental settings for which they have been designed. Our framework develops an SDCA by analyzing the multimodal components of user interactions on the context parameters, from an object centered perspective, as outlined in Section 3. This functionality allows our framework to detect and interpret human activities, their associated behavioral patterns, and the user interaction features in any given setting consisting of any kind of context attributes and environment variables.

4. Video-based systems for activity recognition and analysis, such as [12,19] may have several drawbacks associated with their development, functionalities, and performance metrics. According to [39], video 'presents challenges at almost every stage of the research process.' Some of these are the categorization and transcription of data, the selection of relevant fragments, the selection of camera angle, and the determination of the number of frames. By not using viewer-centered image analysis but by using object centered data directly from the sensors, our proposed framework bypasses all these challenges.

5. Some of the frameworks that have focused on fall detection are dependent on a specific operating system or platform or device. These include the smartphone-based fall detection approach proposed in [23] that uses an Android operating system, the work presented in [29] that uses an IOS operating system, the methodology proposed in [26] that requires a smart cane, and the approach in [15] that requires a handheld device. To address universal diversity and ensure the wide-scale user acceptance of such technologies, it is important that such fall detection systems are platform-independent and can run seamlessly on any device that uses any kind of operating system. Our framework does not have this drawback because it does not need an Android or IOS operating system or any specific device for running. Even though it uses RapidMiner as a software tool to develop its characteristic features, RapidMiner is written in Java—which is platform-independent. RapidMiner allows for the exportation of any process in the form of the associated Java code. Java applications are known as write once run anywhere (WORA). This essentially means that when a Java application is developed and compiled on any system, the Java compiler generates a bytecode or class file that is platform-independent and can be run seamlessly on any other system without re-compilation by using a Java virtual machine (JVM). Additionally, RapidMiner also consists of multiple extensions that can be added to a RapidMiner process and used to seamlessly integrate a RapidMiner process with other applications or software based on the requirements.

6. Several fall detection systems are dependent on external parameters that cannot be controlled and could affect the performance characteristics. For instance, Shao et al. [25] proposed a fall detection methodology based on measuring the vibrations of the floor. Several factors such as the weight of the user, the material of the floor, the condition of the floor, and other objects placed on the floor can impact the intensity of vibrations that could affect the performance of the system. Kong et al.'s [24] system used the distance between the neck of the user and the ground to detect falls. The performance of such a system could be affected by the height of the user, the posture of the user, and any elevations on the ground such as high objects or stairs. The work proposed in [27] by Keaton et al., which used WiFi channel state data to detect falls, could be affected by external factors that tend to influence the WiFi channel state data. Similarly, the methodology developed in [20] worked by using an air pressure sensor, the readings of which could be affected by environmental factors and external phenomena. Such influences or effects of external conditions could have a negative effect on the operational and performance characteristics of the system, and it could even lead to false alarms [40] in caregivers and medical personnel. Such false alarms and alert fatigue can decrease the quality of care, increase response time, and make caregivers and medical personnel insensitive to the warnings of such fall detection systems. The challenge is therefore to ensure that such fall detection systems can seamlessly function without being dependent on external factors that could affect its operation or performance metrics. Our framework uses concepts of complex activity recognition [30] and two related works [31,32], as well as taking the context-driven approach outlined in Section 3, for the analysis of diverse components of user interactions performed on context parameters to interpret the dynamics of human behavior and their relationships with the contextual and spatial features of an environment to detect any anomalies that could constitute an emergency. The performance, operation, and functionality of such an approach is independent of the effect of any external factors or conditions, such as floor vibrations, WiFi channel state data, and the distance between the user and the ground.

## 6. Conclusions and Scope for Future Work

Ambient intelligence in the future of smart homes and smart cities has the potential to address the multiple elderly needs during ADLs due to the behavioral, physical, mental, psychological, and other forms of impairments or limitations that they face with increasing age. A key to developing ambient intelligence in order to address and access these needs lies in monitoring human behavior while analyzing the multimodal components of user interactions with the dynamic contextual, spatial, and temporal features of a given IoT-based ubiquitous environment in which these activities are performed. Therefore, this work provides an interdisciplinary framework that takes a comprehensive approach to study, track, monitor, and analyze human behavior during ADLs. Based on the understanding of the behaviors associated with ADLs, abnormal behaviors leading to situations that might have resulted in health-threatening situations, such as from a fall or unconsciousness, that would need the immediate attention of caregivers or medical practitioners can be detected, and necessary actions can be taken accordingly.

The framework has two novel functionalities that were implemented and tested with an existing dataset. First, it is able to analyze multimodal components of user interactions to identify a list of distinct behavioral patterns associated with each ADL. Using the given dataset, the results showed that it achieved an overall performance accuracy of 76.71%. Second, it uses an intelligent decision-making algorithm that can analyze these behavioral patterns and their relationships with the dynamic contextual and spatial features of the environment to detect any anomalies in user behavior that could constitute an emergency, such as from a fall or unconsciousness. This algorithm achieved an overall performance accuracy of 83.87% when tested on a dataset consisting of multiple ADLs.

To the best of the authors' knowledge, no similar work has been done yet. The presented and discussed results uphold the immense potential and relevance of the framework for the development of ambient intelligence in the future of ubiquitous living environments, e.g., smart homes, to address multiple needs associated with the aging population. Our framework addresses several limitations and challenges in this field, but at this point, its functionality is limited to one user in the confines of a given IoT-based space. Future work along these lines would involve extending the functionality of the framework to incorporate multiple users. We also plan to implement this framework in real-time by setting up an IoT-based environment and incorporating relevant practices and measures to address the healthcare- and safety-related needs of the elderly.

**Author Contributions:** Conceptualization, N.T. and C.Y.H.; methodology, N.T.; software, N.T.; validation, N.T.; formal analysis, N.T.; investigation, N.T.; resources, N.T.; data curation, N.T.; visualization, N.T.; data analysis and results, N.T.; writing—original draft preparation, N.T.; writing—review and editing, C.Y.H. and N.T.; supervision, C.Y.H.; project administration, C.Y.H.; funding acquisition, Not Applicable. All authors have read and agreed to the published version of the manuscript.

**Funding:** This research received no external funding.

**Institutional Review Board Statement:** Not applicable.

**Informed Consent Statement:** Not applicable.

**Data Availability Statement:** Publicly available datasets were analyzed in this study. This data can be found here: https://doi.org/10.17632/sy3kcttdtx.1, accessed on 13 February 2021.

**Conflicts of Interest:** The authors declare no conflict of interest.